\documentclass[aps,pra,twocolumn,english,notitlepage,nofootinbib,floatfix,longbibliography,superscriptaddress]{revtex4-2}
\usepackage[T1]{fontenc}
\usepackage{siunitx}
\usepackage{comment}
\usepackage{physics}
\usepackage{graphicx}
\usepackage{placeins}
\usepackage{textcomp}
\usepackage{amssymb}
\usepackage{amsmath}
\usepackage{makecell}
\usepackage[dvipsnames]{xcolor}
\usepackage[english]{babel}
\usepackage{booktabs}
\usepackage[colorlinks=true,linkcolor=blue,urlcolor=blue,citecolor=blue,pdfusetitle]{hyperref}
\usepackage{tabularx}
\usepackage{array}
\usepackage{booktabs}
\usepackage{soul}

\usepackage{times, txfonts}

\begin{document}

\title{Coherent phase control of two-color continuous variable entangled light}

\author{Andrea Grimaldi}
\affiliation{Niels Bohr Institute, University of Copenhagen, Copenhagen, Denmark}

\author{Valeriy Novikov}
\affiliation{Niels Bohr Institute, University of Copenhagen, Copenhagen, Denmark}

\author{Túlio Brito Brasil}
\email[Contact author: ]{tulio.brasil@nbi.ku.dk}
\affiliation{Niels Bohr Institute, University of Copenhagen, Copenhagen, Denmark}

\author{Eugene Simon Polzik}
\affiliation{Niels Bohr Institute, University of Copenhagen, Copenhagen, Denmark}

\date{\today}

\begin{abstract}
\noindent A continuous variable Einstein-Podolsky-Rosen (EPR) state is a resource for secure quantum communication and distributed quantum sensing. Here we present a technique for coherent control of the two-color EPR state generated by a frequency nondegenerate optical parametric oscillator. The scheme allows for robust control of the homodyne  detection of each of the two EPR quantum fields separated by 200 nanometers. We apply our control scheme to stabilize and characterize a strong entangled state of two-color light displaying 9 dB of two-mode squeezing in the acoustic frequency range, making it a valuable tool for quantum networking and quantum metrology.
\end{abstract}

\maketitle

\section{Introduction}

Hybrid quantum networks require sharing of a quantum state between disparate physical platforms \cite{Kimble2008, Wehner2018}. Therefore, quantum correlations between optical fields at different frequencies are considered one of the most essential elements of such networks. Sources of two-color entanglement in the discrete variable (DV) domain have been used as a quantum channel between distant quantum memories \cite{LagoRivera2021}. However, the continuous variable (CV) counterparts have not achieved the degree of control and correlation strength suitable for such applications. In particular, the deployment of high-quality CV entanglement imposes strict requirements on the phase control, which is still an open challenge for two-color sources.

Here, we propose and demonstrate a technique for the phase stabilization of a highly nondegenerate CV entanglement source. The stabilization scheme enables detection of strong two-mode squeezing, and validation of EPR entanglement criterion at acoustic frequencies. The degree of correlation and the range of sideband frequencies of the two-color EPR state demonstrated in this work make it a valuable resource for distributed quantum sensing.

\begin{figure*}[htp]
\centering
\includegraphics
{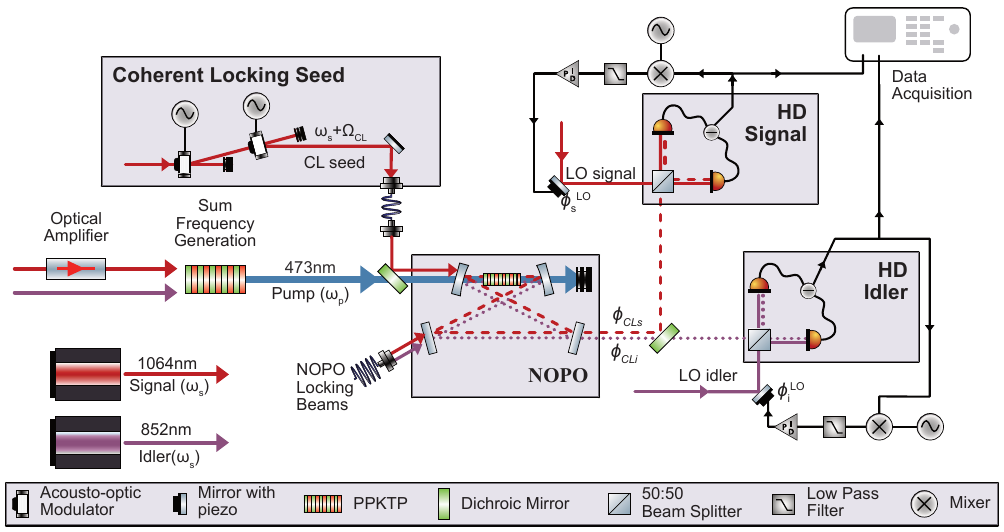}
\caption{Simplified experimental setup for coherent phase control and two‑color EPR state generation. An amplified 1064 nm Nd:YAG laser and a tunable 852 nm Ti:Sapphire laser are combined in a nonlinear crystal to produce a 473 nm pump field ($\omega_p$) via sum‑frequency generation. The pump drives a nondegenerate optical parametric oscillator (NOPO) in a single‑pass configuration, generating entangled vacuum modes at $\omega_s$ and $\omega_i$ resonant at the NOPO cavity (dashed red and purple). A double‑AOM system shifts a small portion of the 1064 nm beam by $\Omega_{\text{CL}}$ to create a coherent control seed, which is injected into the NOPO. Inside the cavity, this seed and the pump produce two bright locking fields, CLs at $\omega_s+\Omega_{\text{CL}}$ (phase $\phi_{\text{CLs}}$) and CLi at $\omega_i -\Omega_{\text{CL}}$ (phase $\phi_{\text{CLi}}$), that co‑propagate with the entangled modes. A dichroic mirror spatially separates signal and idler; each is mixed with its local oscillator at a 50:50 beam splitter. The resulting homodyne photocurrents are demodulated at $\Omega_{\text{CL}}$ to yield phase error signals, which are fed back to piezo‑mounted mirrors in the LO paths to enforce the necessary phase condition with respect to the pump phase. Additional weak counter‑propagating fields at $\omega_s$ and $\omega_i$ provide Pound–Drever–Hall locking of the NOPO cavity for simultaneous resonance of both down‑converted modes.}
\label{fig1:idea scheme}
\end{figure*}

The stabilization scheme introduced here generalizes the coherent control techniques developed for single-mode squeezing \cite{Vahlbruch2006} to the two-mode EPR state of light. As illustrated in Fig.\ref{fig1:idea scheme}, our setup employs a single coherent seed beam, injected at the wavelength of one of the entangled modes into a nondegenerate optical parametric oscillator (NOPO). This seed and its parametrically amplified counterpart at the wavelength
of the other entangled mode provide stable phase references, enabling precise control over the relative phase between the entangled modes. The simplicity and effectiveness of this configuration allow for the observation of strong EPR correlations across a broad frequency range.

\section{Two-color EPR state}

In an NOPO, a nonlinear medium with second-order nonlinear susceptibility $\chi^{(2)}$ is placed inside an optical cavity and driven by a classical pump field, see Fig.~\ref{fig1:idea scheme}. The cavity is engineered to be resonant with two distinct modes generated via parametric down-conversion, denoted by the annihilation operators $\hat{a}_s$ (signal) with frequency $\omega_s$ and $\hat{a}_i$ (idler) with frequency $\omega_i$. The frequencies of these modes satisfy the energy conservation, $\hbar\omega_s + \hbar\omega_i = \hbar\omega_p$, where $\omega_p$ is the pump frequency. The mode operators obey the standard bosonic commutation relations, $[ \hat{a}_j(t), \hat{a}_k^\dagger(t') ] = \delta_{jk}\delta(t-t')$, with $j,k \in \{s,i\}$. The interaction between the pump and the intracavity modes is governed by the effective Hamiltonian
\begin{equation}
\hat{H}_I = i\hbar g \left( A_p e^{i\phi_p} \hat{a}_s^\dagger \hat{a}_i^\dagger - A_p^* e^{-i\phi_p} \hat{a}_s \hat{a}_i \right),
\end{equation}
where $g$ is the nonlinear coupling strength, $\phi_p$ is the pump phase and $A_p = \alpha_p e^{-i\omega_p t}$ represents the classical pump field. This Hamiltonian describes creation (or annihilation) of photon pairs in the signal and idler modes resulting in strong EPR entanglement of the output modes $\hat{b}_s$ and $\hat{b}_i$ emerging from the NOPO \cite{Reid1989, Drummond1990}. The optical cavity bandwidth determines the spectral properties of the EPR entanglement while the pump power, optical losses, classical noise contamination, and phase stability of the system determine the observed correlations strength.

In the standard description of the EPR state the pump phase is a constant, usually set to $\phi_p = 0$, defining a reference frame for the entangled modes \cite{Drummond1990,Schori2002}. One can then  write the quantum fields in terms of their quadrature operators $\hat{x}_k = (\hat{b}_k + \hat{b}_k^\dagger)/\sqrt{2}$ and $\hat{p}_k = (\hat{b}_k - \hat{b}_k^\dagger)/(\sqrt{2}i)$, where $k \in \{s, i\}$ denotes the signal and idler modes, respectively. These quadratures satisfy the canonical commutation relation $[\hat{x}_n(t), \hat{p}_m(t')] = 2i\delta_{nm}\delta(t-t')$ with $m,n \in \{s, i\}$. One of the signatures of the EPR state is the presence of strong quantum correlations that lead to the suppression of fluctuations in the two-mode quadratures $\hat{X} = \hat{x}_s - \hat{x}_i$ and $\hat{P} = \hat{p}_s + \hat{p}_i$, which commute and can thus be simultaneously squeezed. Accessing these correlations experimentally requires precise control over the relative detection phases of the entangled modes with respect to the pump phase. 

In the context of single-mode squeezing, the control of orientation of the noise ellipse in relation to the pump field phase is a well known technical problem that  limits the observed quantum noise reduction, since phase noise fluctuations will project antisqueezed  noise into the detected quadrature \cite{Takeno2007,Oelker2016}. To overcome this problem, Vahlbruch \emph{et al} introduced a coherent control scheme where a bright reference field experiences the same parametric amplification and co-propagates with the squeezed beam to serve as a phase reference for homodyne detection \cite{Vahlbruch2006,Chelkowski2007}.  The scheme relies on the reference field phase-locked to the pump and to the LO via second harmonic generation. It is well known that single-mode squeezing can be seen as entanglement between optical sidebands \cite{Zhang2003,Hage2010,Zippilli2015}. The coherent control technique was used to measure entanglement between two near-degenerate modes  \cite{Schori2002,Yap2020}.

In the absence of active stabilization, the pump phase drifts in relation to the local oscillators . We can adopt a phase-space description that is decoupled from the pump reference frame, allowing the pump phase $\phi_p$ to vary. The nonlinear interaction in the NOPO  guarantees that the input fields $\hat{a}_s e^{i\phi_s}$ and $\hat{a}_i e^{i\phi_i}$ satisfying the condition
\begin{equation}
\phi_{i} = \phi_{p}-\phi_{s},
\label{eq:EntPhasesCondition}
\end{equation}
will be maximally correlated via the nonlinear interaction \cite{Schori2002}. Based on this observation, we define the output modes in a rotating frame as $\hat{b}_s(\phi_s) = \hat{b}_s e^{-i\phi_s}$ and $\hat{b}_i(\phi_i) = \hat{b}_i e^{-i\phi_i}$, and introduce generalized quadrature operators:
\begin{align}
\hat{x}_k(\phi_k) = \frac{1}{\sqrt{2}} \left( \hat{b}_k e^{-i\phi_k} + \hat{b}_k^\dagger e^{i\phi_k} \right),\\
\hat{p}_k(\phi_k) = \frac{1}{\sqrt{2}i} \left( \hat{b}_k e^{-i\phi_k} - \hat{b}_k^\dagger e^{i\phi_k} \right).  
\end{align}
These quadratures obey the standard commutation relations $[\hat{x}_n(\phi_n, t), \hat{p}_m(\phi_m,t')] = 2i \delta_{nm}\delta(t-t')$ with $m,n \in \{s, i\}$. Due to the non-classical correlations characteristic of the EPR state, e.g., the two-mode quadratures $\hat{X}(\phi_s, \phi_i) =  \hat{x}_s(\phi_s) - \hat{x}_i(\phi_i)$ and $\hat{P}(\phi_s, \phi_i) = \hat{p}_s(\phi_s) + \hat{p}_i(\phi_i)$ exhibit reduced quantum noise when the phase condition in Eq.~\eqref{eq:EntPhasesCondition} is satisfied.

\subsection{Coherent reference fields}

In our scheme, we inject into the NOPO a seed field for coherent lock $A_{\text{CL}}=\alpha_{\text{CL}}e^{-i\omega_{\text{CL}}t}$ at frequency $\omega_{\text{CL}} = \omega_{s} + \Omega_{\text{CL}}$, through a high reflective mirror, as illustrated in Fig.~\ref{fig1:idea scheme}. As a result of the nonlinear interaction with the pump inside the NOPO, the seed is parametrically amplified, and simultaneously a second coherent field is produced by difference frequency generation (DFG) with  frequency near the idler mode. We refer to these fields as coherent locking fields CLs with carrier frequencies $\omega_{\text{CLs}} = \omega_{\text{CL}}$ and CLi with carrier frequencies $\omega_{\text{CLi}} = \omega_i - \Omega_{\text{CL}}$, for signal and idler respectively. The phase relation defined in Eq.~\eqref{eq:EntPhasesCondition} is imposed on the coherent locking fields through the nonlinear interaction.

\begin{figure}[htp!]
    \includegraphics[width=0.48\textwidth]{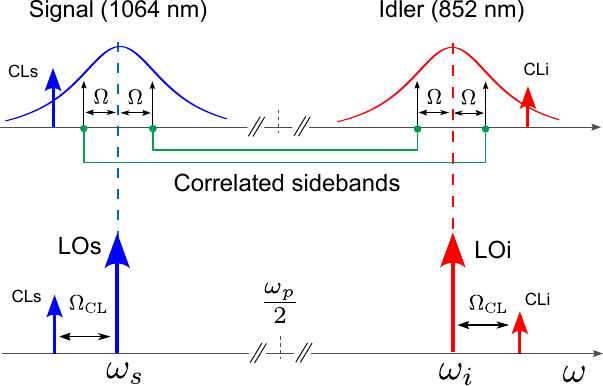}  \\
\caption{Optical spectra of entangled modes, coherent control fields, and local oscillators.
(Top) Signal ($\omega_s$) and idler ($\omega_i$) modes lie symmetrically around the half the pump frequency  $\omega_p/2$. Bright coherent lock fields CLs and CLi are detuned by $\Omega_{\text{CL}}$ and $-\Omega_{\text{CL}}$, respectively, from each cavity resonance — far enough from the EPR sideband band to prevent seed‑noise contamination $\Omega\ll\Omega_{\text{CL}}$, yet within the cavity linewidth for efficient parametric amplification.
(Bottom) During homodyne detection, each local oscillator (LOs, LOi) beats with its corresponding coherent lock. Demodulation of these beat notes provides the error signals used to lock the LO phases to the pump, enforcing the entanglement phase condition.}
\end{figure}

To infer the dynamics of CLs we describe the behavior of  nondegenerate optical parametric amplifier in the classical regime. With injected  seed $A_{\text{CL}}$, the classical intracavity fields obey the following equations of motion

\begin{subequations}
\label{eq:OPOmotionclassic}
\begin{align}
&\dot{\alpha}_{s}=
-(\gamma_s-i\Delta_s)\alpha_s+G\alpha^{\ast}_{i}+\sqrt{2\gamma_s^{in}}\alpha_{\text{CL}} \label{eq:OPOmotionclassic-s} \\
&\dot{\alpha}_{i}=
-(\gamma_i-i\Delta_i)\alpha_i+G\alpha^{\ast}_{s},
\label{eq:OPOmotionclassic-i}
\end{align}
\end{subequations} where $G=g\alpha_p$ is the rescaled nonlinear coupling strength, $\gamma_{s}$ and $\gamma_{i}$ are the total decay rates, $\Delta_s$ and $\Delta_i$ are the cavity detunings, for signal and idler respectively.
For the signal $\gamma_{s}\equiv\gamma_{s}^{in}+\gamma_{k}^{out}+\mu_{k}$ with the rates $\gamma_{s}^{in}$  due to the input mirror for the seed injection, $\gamma_{s}^{out}$ due to the output-coupler and $\mu_{k}$ taking into account spurious intracavity losses. In our case, we can assume $\gamma_{s}=\gamma_{i}=\gamma$ and the output-coupler with same transmission for both wavelengths $\gamma^{\text{out}}_s=\gamma^{\text{out}}_i=\gamma^{\text{out}}$. If the cavity length is stabilized to the resonance for both fundamental modes, then CLs and CLi should have opposite detunings leading to $\Delta_s=-\Delta_i\equiv\Delta$, since these two fields are generated symmetrically with respect to $\omega_{p}/2$.
The detuning $\Delta$ can be selected such that both $\Omega_{\text{CLs}}$ and $\omega_{\text{CLi}}$ are well within the cavity linewidth $\gamma$, while remaining far from the sideband frequencies of the EPR states of interest. This ensures that the locking fields experience the cavity enhanced nonlinear interaction without disturbing the quantum correlations of the entangled modes.

From the steady state solution of Eqs.~\eqref{eq:OPOmotionclassic}, $\dot{\alpha}_{s}=0$ and  $\dot{\alpha}_{i}=0$, we obtain the expressions for the amplitudes of the coherent control fields exiting the cavity

\begin{subequations}\label{eq:OPOsingleseed}
\begin{align}
    A_{\text{CLs}} &= \frac{2\sqrt{\gamma^{in}\gamma^{out}} \gamma^{-1}}{\left(1-\frac{\epsilon^{2}_{}}{1+\Delta'^{2}  }\right) -i\Delta'^{} \left(1+ \frac{\epsilon^{2}_{}}{1+\Delta'} \right) }\,\alpha^{}_{\text{CL}}, 
\label{eq:OPOsingleseed_CFL1}\\ 
    A^{}_{\text{CLi}}&=\frac{\epsilon\,e^{i\phi_{p}}}{1-i\Delta' }
A_{\text{CLs}}^{*},
\label{eq:OPOsingleseed_CFL2}
\end{align}
\end{subequations}
where $\Delta'= \Delta / \gamma$ is the normalized detuning and the parameter $\epsilon = \alpha_p / \alpha_{\text{th}}$ is the ratio of the intracavity pump amplitude to the pump threshold amplitude $\alpha_{\text{th}}$. The phases of the coherent locking fields are given by
\begin{subequations}\label{eq:PhaseCLFs}
\begin{align}
\phi_{\text{CLs}} &= \arg\{A_{\text{CLs}}\} \label{eq:PhaseCLF1} \\
\phi_{\text{CLi}} &= \phi_p - \phi_{\text{CLs}}. \label{eq:PhaseCLF2}
\end{align}
\end{subequations}
The phases of the coherent lock fields satisfy the entanglement phase condition of Eq.~\eqref{eq:EntPhasesCondition}, making them suitable phase references for homodyne detection. They enable the EPR correlations to be accessed and measured even in a highly nondegenerate regime. In this analysis, we neglect constant phase shifts introduced by the detuning, as they do not affect the relative phase condition.

\subsection{Phase error signals}

The fields $A_{\text{CLs}}$ and $A_{\text{CLi}}$ co-propagating   with the entangled beams Fig.~\ref{fig1:idea scheme} serve as bright detection phase references. At the homodyne detectors, the local oscillator fields defined as $A^{\text{LO}}_{s} = \alpha^{\text{LO}}_{s} e^{i(\omega_s t + \phi^{\text{LO}}_{s})}$ and $A^{\text{LO}}_{i} = \alpha^\text{LO}_{i} e^{i(\omega_i t + \phi^{\text{LO}}_{i})}$, for signal and idler respectively, interfere with the references producing optical beat notes at frequencies $+\Omega_{\text{CL}}$ for the signal and $-\Omega_{\text{CL}}$ for the idler. The homodyne photocurrents carry information on the relative phases $\theta_s = \phi_{\text{CLs}} - \phi_{s}^{\text{LO}}$ and $\theta_i = \phi_{\text{CLi}} - \phi_{i}^{\text{LO}}$ which can be extracted using standard phase detection techniques. Demodulating the photocurrents gives error signals of the form
\begin{subequations}
\begin{align}
S_{s} &= |\alpha_{s}^{\text{LO}}||A_{\text{CLs}}| \sin(\theta_{s} + \theta_{s}^{\text{ref}})\\
S_{i} &= |\alpha_{i}^{\text{LO}}||A_{\text{CLi}} |\sin(\theta_{i} - \theta_{i}^{\text{ref}}),
\end{align}
\end{subequations}
where $\theta_{s}^{\text{ref}}$ and $\theta_{i}^{\text{ref}}$ are electronic phase references for the phase-locking loops that can be adjusted to change the lock set point. The opposite signs in the arguments of the sine functions reflect the opposite signs of the beat note frequencies for the signal and idler channels. These error signals are used to stabilize the LO phases such that
\begin{subequations}\label{eq:lock_condition}
\begin{align}
\phi_{s}^{\text{LO}} &= \phi_{\text{CLs}} - \theta_{s}^{\text{ref}}, \\
\phi_{i}^{\text{LO}} &= \phi_p - \phi_{\text{CLi}} + \theta_{i}^{\text{ref}},
\end{align}
\end{subequations}
ensuring that the LO phases dynamically follow the coherent control field phases. Under these locking conditions, the homodyne detectors measure the slowly varying envelope quadratures 
\begin{equation}
    \hat{q}_k(\phi_{k}^{\text{LO}}) = \frac{\hat{b}_k e^{-i\phi_{k}^{\text{LO}}} + \hat{b}_k^\dagger e^{i\phi_{k}^{\text{LO}}}}{\sqrt{2}},
\end{equation}
which correspond to linear combinations of the generalized quadratures $\{\hat{x}_k(\phi_k), \hat{p}_k(\phi_k)\}$. Specifically, the measured quadratures are
\begin{subequations}
\begin{align}
\hat{q}_s(\theta_{s}^{\text{ref}}) &= \hat{x}_s(\phi_s) \cos(\theta_{s}^{\text{ref}}) + \hat{p}_s(\phi_s) \sin(\theta_{s}^{\text{ref}}), \\
\hat{q}_i(-\theta_{i}^{\text{ref}}) &= \hat{x}_i(\phi_i) \cos(\theta_{i}^{\text{ref}}) + \hat{p}_i(\phi_i) \sin(\theta_{i}^{\text{ref}}),
\end{align}
\end{subequations}
where the quadrature angles are determined by the setpoints $\theta_{s}^{\text{ref}}$ and $\theta_{i}^{\text{ref}}$ of the respective phase-locking loops, but with opposite sign.

\subsection{Two-mode squeezing and entanglement}

The condition for observing maximum quantum correlations between the measured quadratures $\hat{q}_s(\theta_{s}^{\text{ref}})$ and $\hat{q}_i(\theta_{i}^{\text{ref}})$ is satisfied when the detection angles are equal, i.e., $\theta_{s}^{\text{ref}} = \theta_{i}^{\text{ref}} \equiv \theta^{\text{ref}}$. Under this condition, the detection phases imposed by Eq.~\eqref{eq:lock_condition} align with the entanglement phase condition in Eq.~\eqref{eq:EntPhasesCondition}, enabling the observation of quantum noise suppression in the two-mode observable 
\begin{equation}
    \hat{Q}_{-} = \frac{\hat{q}_s(\theta_{s}^{\text{ref}}) - \hat{q}_i(\theta_{i}^{\text{ref}}) }{\sqrt{2}},
\end{equation}
and a corresponding noise enhancement in the orthogonal combination 
\begin{equation}
\hat{Q}_{+} = \frac{\hat{q}_s(\theta_{s}^{\text{ref}}) + \hat{q}_i(\theta_{i}^{\text{ref}})}{\sqrt{2}},
\end{equation}
as discussed in Ref.~\cite{Schori2002}. 
\begin{figure}[htp!]
    \includegraphics[width=0.48\textwidth]{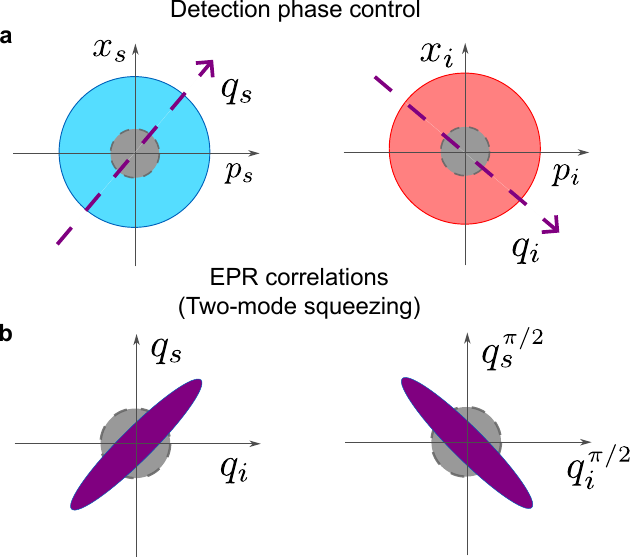}  \\
\caption{Control of simultaneous detection of each entangled modes and and two-mode squeezing. (a) Phase space for each entangled mode. From the parametric amplification the individual states of signal and idler present excess of noise (thermal state), larger than the single-mode vacuum noise (gray circle). The purple arrows represent the selected quadrature projections defined by the local oscillators when the right phase condition in relation to the pump phase is fulfilled. (b) EPR correlations seen as two-mode squeezing. The measurement of $q_s$ and $q_i$ are correlated, leading to noise bellow the two-mode vacuum noise (enlarged gray circle), the measurement of the conjugated quadratures is anti-correlated.}
\end{figure}
Assuming identical detection efficiency $\eta$ and decay rates $\gamma$ for the signal and idler modes of the NOPO cavity, the noise spectral density of these observables at Fourier frequency $\Omega$ is given by 
\begin{equation}
    \Delta Q_{\pm}^2(\Omega) = 1 \pm \eta \frac{4\epsilon}{\Omega^2/\gamma^2 + (\epsilon + 1)^2},
    \label{eq:tm_sqz_asqz}
\end{equation}
where $\delta(\Omega-\Omega')\Delta Q_{\pm}^2(\Omega) = \langle \hat{Q}_{\pm}(\Omega) \hat{Q}_{\pm}(-\Omega') \rangle$, and $\epsilon = \alpha_p / \alpha_{p,\text{th}}$ is the normalized pump amplitude. A similar behavior is observed for the orthogonal quadratures $\hat{q}_s(\theta_{s}^{\text{ref}} + \pi/2)$ and $\hat{q}_i(\theta_{i}^{\text{ref}} - \pi/2)$, which correspond to measurements of the $\hat{q}_s(\theta_{s}^{\text{ref}})$ conjugate quadrature. In this case, the joint observables 
\begin{equation}
   \hat{Q}_{\pm}^{\pi/2} =  \frac{\hat{q}_s(\theta_{s}^{\text{ref}} + \pi/2) \pm \hat{q}_i(\theta_{i}^{\text{ref}} - \pi/2)}{\sqrt{2}},
\end{equation}
exhibit the complementary noise behavior $\Delta (Q_{\pm}^{\pi/2})^2 = \Delta Q_{\mp}^2$. For nonidentical detection efficiencies see Ref~\cite{Schori2002}.

Well known criteria for detecting quantum entanglement such as that of Duan-Simon \cite{Duan2000,Simon2000}, are based on the measurement of the variances $\langle \Delta(\hat{x}_s - \hat{x}_i)^2 \rangle$ and $\langle \Delta(\hat{p}_s + \hat{p}_i)^2 \rangle$. These criteria can be naturally extended to the generalized quadratures $\hat{q}_s(\theta^{\text{ref}})$, $\hat{q}_i(\theta^{\text{ref}})$, and their orthogonal counterparts $\hat{q}_s(\theta^{\text{ref}} + \pi/2)$, $\hat{q}_i(\theta^{\text{ref}} - \pi/2)$, which are accessible through phase-controlled homodyne detection. For instance, the Duan-Simon inseparability criterion says that for any separable state 
\begin{equation}
    \Delta (Q_{+}^{\pi/2})^2 + \Delta Q_{-}^2 \geq 2.
\end{equation} 
Violation of this inequality provides a sufficient condition to demonstrate entanglement between the two modes.

\section{The impact of phase noise}

Deviations from the phase conditions imposed by Eq.~\eqref{eq:lock_condition} result in a mismatch with the entanglement phase condition of Eq.~\eqref{eq:EntPhasesCondition}, leading to degradation of the observed quantum correlations. In practice, such deviations arise from residual phase noise in the locking loops, which can be modeled as small stochastic fluctuations around the optimal locking points.

Let $\delta\theta_k$ denote the Gaussian-distributed phase noise affecting the relative phase between the entangled mode $k \in \{s, i\}$ and its corresponding local oscillator. These fluctuations are centered around the ideal setpoint $\bar{\theta}_{k}^{\text{ref}}$, such that the actual detection phase becomes $\theta_k = \bar{\theta}_{k}^{\text{ref}} + \delta\theta_k$. To analyze the effect of phase noise on the joint quadrature observables $\hat{Q}_{\pm}$, we decompose the fluctuations into $(\delta\theta_s + \delta\theta_i)/2 $ (common mode) and $(\delta\theta_s - \delta\theta_i)/2$ (differential mode). Inspection of Eq.~\eqref{eq:lock_condition} reveals that only the common-mode fluctuation $\Theta \equiv (\delta\theta_s + \delta\theta_i)/2 $ affects the measurement of $\hat{Q}_{\pm}$, while the differential mode can be absorbed into a redefinition of the seed phase, $\phi_{\text{CLs}}' = \phi_{\text{CLs}} + (\delta\theta_s - \delta\theta_i)/2$. Without loss of generality, we therefore set $(\delta\theta_s - \delta\theta_i)/2 = 0$ and consider the case where both detection phases fluctuate symmetrically around a common setpoint. 

Under these assumptions, the measured state becomes a mixture of rotated quadratures
\begin{equation}
\Delta Q_{\pm,\text{pn}}^2 = \int \left(\Delta Q_{\pm}^2 \cos^2\Theta + \Delta (Q_{\pm}^{\pi/2})^2 \sin^2 \Theta\right) P(\Theta) \, d\Theta,
\end{equation}
where \(P(\Theta)\) is a zero-mean Gaussian distribution with the variance 
$\sigma_\Theta^2 =  (\sigma_{\delta\theta_s}^2 + \sigma_{\delta\theta_i}^2 + 2\,\text{Cov}(\delta\theta_s, \delta\theta_i) )/4$ . The presence of phase noise causes mixing between the squeezed quadrature $\hat{Q}_{\pm}$ and its orthogonal counterpart $\hat{Q}_{\pm}^{\pi/2}$, resulting in a degradation of the observed squeezing. Assuming small fluctuations ($\sigma_\Theta \ll 1$), the variance of the measured in the presence of phase noise quadrature becomes
\begin{equation}
\Delta Q_{\pm,\text{pn}}^2 \simeq \Delta Q_{\pm}^2 (1 - \sigma_\Theta^2) + \Delta (Q_{\pm}^{\pi/2})^2 \sigma_\Theta^2,
\label{eq:TMSQZ_phasenoise}
\end{equation}
where \(\Delta Q_{\pm}^2\) and \(\Delta (Q_{\pm}^{\pi/2})^2\) are the variances of the ideal squeezed and anti-squeezed quadratures, respectively. The expression Eq~\eqref{eq:TMSQZ_phasenoise} quantifies the sensitivity of two-mode squeezing measurements to residual phase noise and highlights the importance of precise phase stabilization. Note that the requirement on the size of the phase fluctuations $\sigma_{\Theta}$ becomes more stringent for higher degrees of squeezing. 

\section{Experimental setup and results}

The proposed coherent control technique is implemented and validated using a two-color entanglement source operating at wavelengths $\lambda_s = 1064$ nm and $\lambda_i = 852$ nm. These modes are separated by approximately 35 THz from the system’s central frequency $\omega_p/2$, making conventional locking strategies~\cite{Yap2020} technically challenging. 

\begin{figure}[htp!]
    \includegraphics[width=0.5\textwidth]{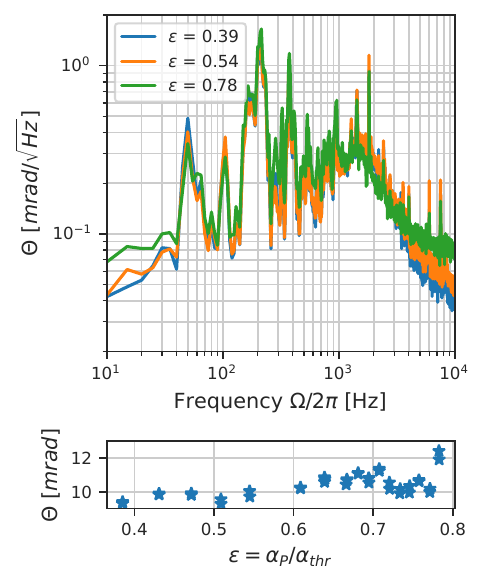}  \\
\caption{Estimation of phase noise via coherent control error signals.
(Top) Power spectral density (PSD) of the common‑mode phase fluctuation $\Theta = (\delta\theta_s + \delta\theta_i)/2$ plotted against analysis frequency for three normalized pump amplitudes: $\epsilon = 0.39$ (blue), 0.54 (orange), and 0.78 (green). The PSDs are calibrated by scanning each local‑oscillator phase and measuring the resulting peak‑to‑peak error‑signal amplitudes. (Bottom) Root‑mean‑square phase noise $\sigma_{\Theta}$, obtained by integrating PSD over the detection band, shown as a function of $\epsilon$. The roughly constant $\sigma_{\Theta} = 10 \pm 2$ mrad highlights the stability and robustness of the locking scheme.}
\label{fig:PhaseNoise}
\end{figure}

The NOPO (Fig.~\ref{fig1:idea scheme}) is constructed using a type-0 periodically poled KTP crystal placed inside a bow-tie cavity with a free spectral range (FSR)  $\approx$ 670 MHz and a linewidth of $\gamma = 15$ MHz for both signal and idler modes. Double resonance at the entangled wavelengths is maintained using a two-stage Pound-Drever-Hall (PDH) locking technique. A key feature of the setup is the use of two independent lasers: a Ti:Sapphire laser at 852 nm and a Nd:YAG laser at 1064 nm. These beams are combined via sum-frequency generation to produce the pump field at $\lambda_p = 473$ nm. Portions of the laser outputs are spatially filtered using mode-cleaning cavities and used as local oscillators for homodyne detection. The same lasers also provide weak coherent fields for PDH locking of the NOPO cavity. Further details on the cavity design and stabilization can be found in Ref.~\cite{Brasil2022}.

During the experiment, the NOPO was pumped with up to 180 mW at 473 nm, corresponding to a normalized pump amplitude $\epsilon = 0.75$, yielding a parametric gain $G \lesssim 4$. The generalized coherent control scheme was implemented by injecting a single seed field at $\lambda_s = 1064$ nm. The seed was detuned by $\Omega_s/2\pi \approx 3$ MHz — well within the cavity linewidth but outside the bandwidth of interest for our entanglement measurements. The detuning was generated using two acousto-optic modulators (AOMs) driven at $\nu_1 = 78$ MHz and $\nu_2 = 75$ MHz, producing a net shift of $\nu_1 - \nu_2 = 3$ MHz. Both AOMs were driven by direct digital synthesizers (DDS) synchronized to a common clock reference. A 10 $\mu$W seed was injected into the NOPO through the same port as the pump. After transmission through the cavity, the resulting CLs and CLi fields were superimposed with the corresponding local oscillators giving a homodyne detection clearance of $24$ dB over the detector dark noise.

For both beams the photocurrent from the homodyne detection is sent to a SR830 lock-in amplifier to detect the beat note between the coherent lock field and local oscillator. Demodulation was performed using DDS-generated reference tones at 3 MHz, synchronized with the AOM drivers to eliminate frequency drift. The resulting error signals were fed back to piezoelectric transducers (PZTs) mounted on mirrors in the LO paths to stabilize the optical phase.

\begin{figure}[htp!]
\centering
\includegraphics[width=0.48\textwidth]{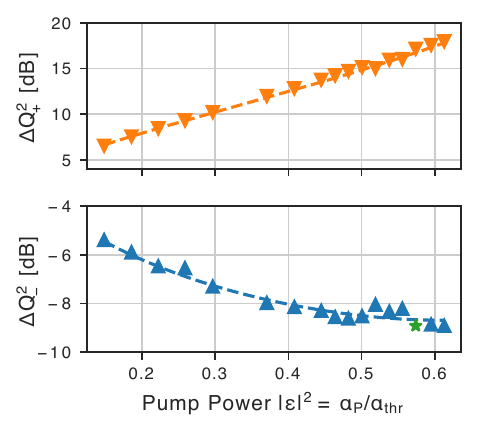}
\caption{Two‑mode squeezing and anti‑squeezing versus normalized pump power. Measured noise variances $\Delta Q^{2}_{-}$ (squeezing, blue triangles) and $\Delta Q^{2}_{+}$ (anti‑squeezing, orange triangles) are plotted as functions of the normalized pump power, with each data point representing the RMS noise in the 5–15 kHz band after optimally combining the two homodyne photocurrents and locking the phases. Solid curves are fits to the phase‑noise model in Eq.~\eqref{eq:TMSQZ_phasenoise}, yielding an overall detection efficiency $\eta = 0.89 \pm 0.01$ and a residual common‑mode phase noise $\tilde{\sigma}_\Theta = 8 \pm 5 $ mrad. The maximum two‑mode squeezing (green star) of 9 dB is achieved near $\epsilon= 0.8$, demonstrating the robustness of the coherent‑control scheme.}
\label{fig:entresults}
\end{figure}

These coherent control error signals were also used to validate the phase noise model described by Eq.~\eqref{eq:TMSQZ_phasenoise}. The error signals were calibrated by scanning the LO phases and measuring the peak-to-peak amplitudes $S_{\text{pp},k}$, yielding calibration factors $\beta_k = 2/S_{\text{pp},k}$. Using these data, we estimated the total phase noise $\sigma_\Theta$ for each squeezing measurement. Fig.~\ref{fig:PhaseNoise} shows the power spectral densities (PSDs) of $\Theta = (\delta\theta_s + \delta\theta_i)/2$ and total phase noise for different squeezing measurements, performed under varying pump powers and setpoints $\theta^{\text{ref}}_s$ and $\theta^{\text{ref}}_i$. The spectra exhibit consistent features, indicating excellent stability of the locking system. From these measurements, we estimate an average phase noise of $\sigma_\Theta = 10 \pm 2$ mrad.

To further validate the error signal calibration, we independently estimated the phase noise by analyzing the fluctuations of the combined homodyne photocurrents as a function of the normalized pump amplitude. For each value of $\epsilon$, the detection phases were locked, and the photocurrent time traces $q_s(\theta_{s}^{\text{ref}})$ and $q_i(\theta_{i}^{\text{ref}})$ were recorded. These signals were then combined with a relative weight $g$ to compensate for small unbalance in the detection efficiency of the two EPR beams
\begin{equation}
    Q_{\pm}(g) = \frac{q_s(\theta_{s}^{\text{ref}}) \pm g\,
q_i(\theta_{i}^{\text{ref}})}{\sqrt{2}}.
\end{equation}
Both the relative weight $g$ and the set point $\theta_{s}^{\text{ref}}$ for the signal CL were optimized to maximize two-mode correlations leading to minimum $\Delta Q^{2}_{-}$. The normalized noise variances, $\Delta Q^{2}_{\pm}$, were calculated as the root-mean-square values within the 5-15 kHz bandwidth and normalized to the shot noise level. By interpolating $\Delta Q^{2}_{+}$ and $\Delta Q^{2}_{-}$ using Eq.~\eqref{eq:TMSQZ_phasenoise}, we estimated a total phase noise of $\tilde{\sigma}_\Theta = 8 \pm 5$ mrad, in good agreement with the values inferred from the error signal analysis.

\begin{figure*}[ht!]
\includegraphics[width=1.\textwidth]{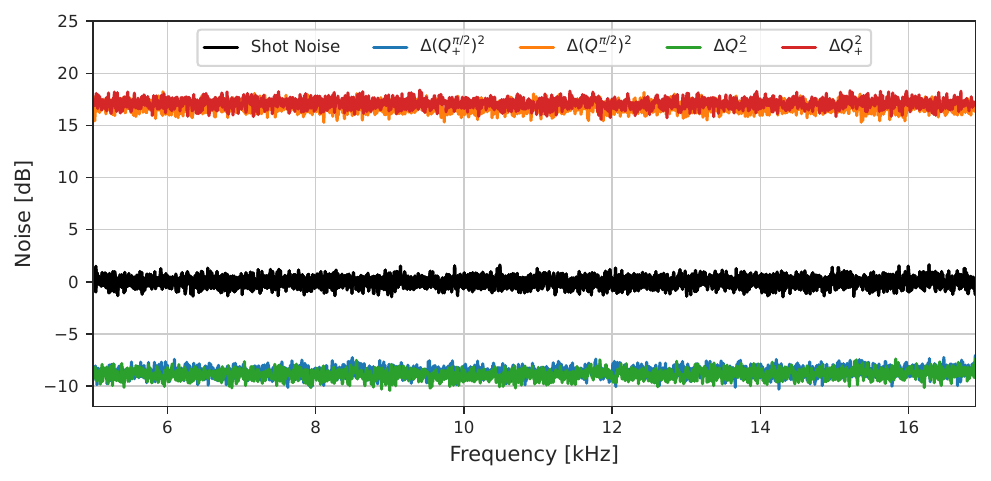}
\caption{EPR-variables noise spectra between 5-17 kHz. All spectra are normalized to the two-mode shot noise level, with the pump amplitude set to $\epsilon \simeq 0.8$. The variances $\Delta Q^{2}_{+}$ and $\Delta (Q^{\pi/2}_-)^{2}$ present 17 dB of anti-squeezing while the variances $\Delta Q^{2}_{-}$ and $\Delta (Q^{\pi/2}_-)^{2}$ reach near 9 dB of two-mode squeezing.}
\label{fig:two_mode_entanglement_spectra}
\end{figure*}

Real-time monitoring of phase noise allows us to optimize the feedback loops and enhance overall system performance. In case of symmetric efficiencies the value of the optimum pump power $\epsilon_{\text{opt}} = \min_{\epsilon}(\Delta Q^{2}{\pm})$ is determined by $\sigma_{\Theta}$. The suppression of phase noise enabled an increase in pump power without compromising the quantum correlations, which now reach their maximum at $\epsilon \simeq 0.8$, significantly improving upon the previously reported value of $\epsilon_{\text{opt}} \simeq 0.5$~\cite{Brasil2022}. This corresponds to the reduction in the phase noise by more than a factor of 10, allowing us to observe up to $9$ dB of two-mode squeezing, thus reducing the performance gap between the state-of-art single-mode \cite{Vahlbruch2016} and two-color two-mode squeezing sources.

To confirm the presence of entanglement we measure $\Delta Q^{2}_{-}$ and $\Delta (Q_+^{\pi/2})^{2}$. Fig~\ref{fig:two_mode_entanglement_spectra} shows the noise spectra in acoustic frequency range 5-17 kHz demonstrating strong two-mode squeezing in both combinations of quadratures. Continuous-variable EPR entanglement is verified by the Duan-Simon criterion giving $\Delta Q^{2}_{-}+\Delta (Q^{\pi/2}_+)^{2} = 0.26(1) \ll 2$.

The long-term stability of the locking system was limited by environmental factors. The AOM-based seed generation setup was connected to the NOPO via polarization-maintaining single-mode fibers, which are sensitive to temperature fluctuations. These induced phase drifts between the signal LO and the seed eventually saturated the PZT actuator after a few minutes. Additionally, the bandwidth of the coherent control loop was constrained by mechanical resonances of the PZT mounts.

\section{Conclusion and Outlook}

We have demonstrated a robust and flexible scheme for phase stabilization of a two-color continuous-variable entanglement source, based on a nondegenerate optical parametric oscillator. A central innovation of our approach is the use of a single coherent seed field injected into the NOPO, which enables generation of coherent locking fields without requiring phase stabilization between the seed and the pump. This marks a significant departure from previous strategies~\cite{Chelkowski2007,Yap2020}, where tight phase locking to the pump was essential.

By decoupling the quadrature phase space from the pump reference frame~\cite{Schori2002}, our method allows for independent and tunable access to the generalized quadratures of each entangled mode. This flexibility is particularly advantageous in hybrid quantum networks and sensing applications, where the entangled modes may operate at widely separated frequencies. We experimentally verified that this generalized phase-space framework does not compromise the ability to detect and quantify entanglement~\cite{CoutinhoDosSantos2005}.

Our phase control technique is robust across a broad range of operating conditions. We reach phase noise levels as low as $\sigma_\Theta = 10 \pm 2$ mrad. The suppression of phase noise effectively eliminated one of the two main sources of squeezing degradation, with the remaining limitation from the average detection efficiency. The simplicity and effectiveness of our coherent control scheme, combined with the ability to achieve squeezing levels exceeding 9 dB, make our EPR source a promising tool for a wide range of applications, including quantum-enhanced metrology~\cite{Barbieri2022}.

\begin{acknowledgments}
The authors acknowledge the supported by VILLUM FONDEN under a Villum Investigator Grant, grant no.\ 25880, by the Novo Nordisk Foundation through Copenhagen Center for Biomedical Quantum Sensing, grant number NNF24SA0088433 and through ‘Quantum for Life’ Center, grant NNF20OC0059939, by the European Union's Horizon 2020 research and innovation program under the Marie Sklodowska-Curie grant agreements No.\ 125101 `EPROXY', and the European Research Council Advanced grant QUANTUM-N. 
\end{acknowledgments}



\clearpage
\onecolumngrid

\end{document}